%% file: main.tex
\documentclass[manuscript]{acmart}
\AtBeginDocument{%
  }

\setcopyright{acmlicensed}
\copyrightyear{2025}
\acmYear{2025}
\acmDOI{XXXXXXX.XXXXXXX}

\usepackage{tikz}
\usepackage{adjustbox}
\usepackage{graphicx}
\usepackage{color}
\usepackage{caption}
\usepackage{subcaption}
\usepackage{multirow}
\usepackage{listings}
\usepackage{url}
\usetikzlibrary{positioning,chains,quantikz2}

\definecolor{codebrown}{rgb}{0.6,0.3,0.0}
\definecolor{lightgray}{rgb}{0.95, 0.95, 0.95}
\definecolor{darkgray}{rgb}{0.639, 0.639, 0.639}
\definecolor{dkgreen}{rgb}{0,0.6,0}
\definecolor{mauve}{rgb}{0.58,0,0.82}
\definecolor{codegreen}{rgb}{0,0.6,0}
\definecolor{codeblue}{rgb}{0,0,1}
\definecolor{coderose}{rgb}{0.8,0,0.5}

\lstdefinestyle{mypython}{
    basicstyle=\ttfamily\footnotesize,
    numbers=left,
    numbersep=-5pt,
    breaklines=true,
    showspaces=false,
    showstringspaces=false,
    tabsize=8,
    language=Python,
    commentstyle=\itshape\color{codebrown},
    frame=single,
	aboveskip=1mm,
	belowskip=1mm,
    xleftmargin=.1\textwidth,
    xrightmargin=.1\textwidth,
	framesep=6pt,
	showstringspaces=false,
	columns=flexible,
	keywordstyle=\color{codeblue},
    keywordstyle=[2]\color{coderose},         
    keywordstyle=[3]\color{codebrown},
	commentstyle=\color{dkgreen},
	stringstyle=\color{mauve},
    rulecolor=\color{darkgray},
	backgroundcolor = \color{lightgray},
    morekeywords=[2]{in,for, with, CunqaCircuit, phase, run, get_QPUs, gather, run_distributed, rcontrol},
    morekeywords=[3]{x, h, swap,cp, crz, measure_and_send, measure_all, expose, remote_c_if, qsend, qrecv}
}

\lstdefinestyle{onelinepython}{
    basicstyle=\ttfamily\footnotesize,
    numbersep=-5pt,
    breaklines=true,
    showspaces=false,
    showstringspaces=false,
    tabsize=8,
    language=Python,
    commentstyle=\itshape\color{codebrown},
    frame=single,
	aboveskip=3mm,
	belowskip=3mm,
    xleftmargin=.1\textwidth,
    xrightmargin=.1\textwidth,
	framesep=6pt,
	showstringspaces=false,
	columns=flexible,
	keywordstyle=\color{codeblue},
    keywordstyle=[2]\color{coderose},         
    keywordstyle=[3]\color{codebrown},
	commentstyle=\color{dkgreen},
	stringstyle=\color{mauve},
    rulecolor=\color{darkgray},
	backgroundcolor = \color{lightgray},
    morekeywords=[2]{in, for, with, CunqaCircuit, phase, run, get_QPUs, gather, run_distributed, rcontrol},
    morekeywords=[3]{x, h, swap,cp, crz, measure_and_send, measure_all, expose, remote_c_if, qsend, qrecv}
}

\lstdefinestyle{mybash}{
  language=bash,
  basicstyle=\ttfamily\small,
  keywordstyle=\color{blue}\bfseries,
  stringstyle=\color{orange!80!black},
  commentstyle=\color{gray}\itshape,
  showstringspaces=false,
  rulecolor=\color{darkgray},
  breaklines=true,
  postbreak=\mbox{\textcolor{gray}{$\hookrightarrow$}\space},
  literate={\$ }{{\textcolor{green!60!black}{\$ }}}1,
  columns=fullflexible
}

\begin{document}

\title{CUNQA: a Distributed Quantum Computing emulator for HPC}

\author{Jorge V\'azquez-P\'erez}
\email{jvazquez@cesga.es}
\orcid{0009-0002-1442-4181}
\author{Daniel Exp\'osito-Patiño}
\orcid{0009-0007-8376-4733}
\email{dexposito@cesga.es}
\author{Marta Losada}
\email{mlosada@cesga.es}
\orcid{0009-0005-3153-0578}
\author{\'Alvaro Carballido}
\email{acarballido@cesga.es}
\orcid{0009-0005-9570-2420}
\author{Andr\'es G\'omez}
\email{agomez@cesga.com}
\orcid{0000-0001-7272-8488}
\affiliation{%
  \institution{Galicia Supercomputing Center (CESGA)}
  \city{Santiago de Compostela}
  \country{Spain}
}
\author{Tom\'as F. Pena}
\email{tf.pena@usc.es}
\orcid{0000-0002-7622-4698}
\affiliation{%
  \institution{Universidad de Santiago de Compostela}
  \city{Santiago de Compostela}
  \country{Spain}
}

\renewcommand{\shortauthors}{V\'azquez-P\'erez et al.}

\begin{abstract}
The challenge of scaling quantum computers to gain computational power is expected to lead to architectures with multiple connected quantum processing units (QPUs), commonly referred to as Distributed Quantum Computing (DQC). In parallel, there is a growing momentum toward treating quantum computers as accelerators, integrating them into the heterogeneous architectures of high-performance computing (HPC) environments. This work combines these two foreseeable futures in CUNQA, an open-source DQC emulator designed for HPC environments that allows testing, evaluating and studying DQC in HPC before it even becomes real. It implements the three DQC models of no-communication, classical-communication and quantum-communication; which will be examined in this work. Addressing programming considerations, explaining emulation and simulation details, and delving into the specifics of the implementation will be part of the effort. The well-known Quantum Phase Estimation (QPE) algorithm is used to demonstrate and analyze the emulation of the models. To the best of our knowledge, CUNQA is the first tool designed to emulate the three DQC schemes in an HPC environment.
\end{abstract}



\received{20 February 2007}
\received[revised]{12 March 2009}
\received[accepted]{5 June 2009}

\maketitle

\input{capitulos/1_introduction}
\input{capitulos/2_cunqa}
\input{capitulos/3_display}

\input{capitulos/4_related_work}
\input{capitulos/5_future_work}
\input{capitulos/6_conclusions}

\input{capitulos/7_data_availability}

\begin{acks}

This work has been mainly funded by the project QuantumSpain, financed by the Ministerio de Transformación Digital y Función Pública of Spain's Government through the project call QUANTUM ENIA – Quantum Spain project, and by the European Union through the Plan de Recuperación, Transformación y Resiliencia – NextGenerationEU within the framework of the Agenda España Digital 2026. J. V\'azquez-P\'erez was supported by the \emph{Axencia Galega de Innovaci\'on (Xunta de Galicia)} through the \emph{Programa de axudas \'a etapa predoutoral (ED481A \& IN606A)}.

Additionally, this research project was made possible through the access granted by the Galician Supercomputing Center (CESGA) to two key parts of its infrastructure. Firstly, its Qmio quantum computing infrastructure with funding from the European Union, through the Operational Programme Galicia 2014-2020 of ERDF\_REACT EU, as part of the European Union’s response to the COVID-19 pandemic.

Secondly, The supercomputer FinisTerrae III and its permanent data storage system, which have been funded by the NextGeneration EU 2021 Recovery, Transformation and Resilience Plan, ICT2021-006904, and also from the Pluriregional Operational Programme of Spain 2014-2020 of the European Regional Development Fund (ERDF), ICTS-2019-02-CESGA-3, and from the State Programme for the Promotion of Scientific and Technical Research of Excellence of the State Plan for Scientific and Technical Research and Innovation 2013-2016 State subprogramme for scientific and technical infrastructures and equipment of ERDF, CESG15-DE-3114.

\end{acks}

\bibliographystyle{ACM-Reference-Format}
\bibliography{biblio}

\end{document}

%% file: capitulos/1_introduction.tex
\section{Introduction}
\label{sec:intro}

Quantum computing is a rather unpredictable and fast-changing field of research, both due to its immaturity---intrinsic to its current stage of development---and the fact that the field is dominated by large private companies in a race to be the first to achieve quantum advantage or, even, supremacy. However, one can analyze the trends driving the field through the currents of most scientific work and by the promises and forecasts issued by the very corporations that own the technology. There are two topics of growing interest that are worth mentioning in relation to this work: first, the integration of quantum processing units (QPUs) into high-performance computing (HPC) environments, envisioning them as GPU-like accelerators; second, the development of a multicore structure in quantum computing, analogous to that found in classical computing, which typically falls under the umbrella of so-called distributed quantum computing (DQC). In fact, the multicore architecture is supported by the roadmaps of several companies (IBM~\cite{IBM_roadmap} or Google~\cite{Google_roadmap}, for example), where they hint at the necessity of such architectures. This work takes these two perspectives as its starting point and aims to bring the two domains together. To do so, it is essential to establish a distinct context for each in order to clarify the logic for their integration.

\begin{figure*}[t!]
    \centering
    \begin{subfigure}[t]{0.3\textwidth}
        \centering
        \includegraphics[height=1.7in]{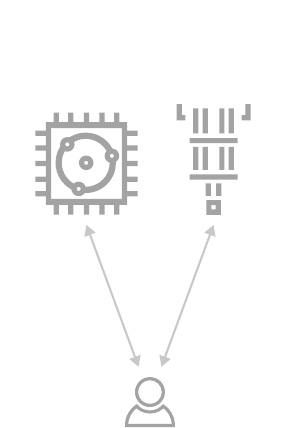}
        \caption{Standalone.}
        \label{subfig:hpcqc_integration_a}
    \end{subfigure}%
    ~ 
    \begin{subfigure}[t]{0.3\textwidth}
        \centering
        \includegraphics[height=1.7in]{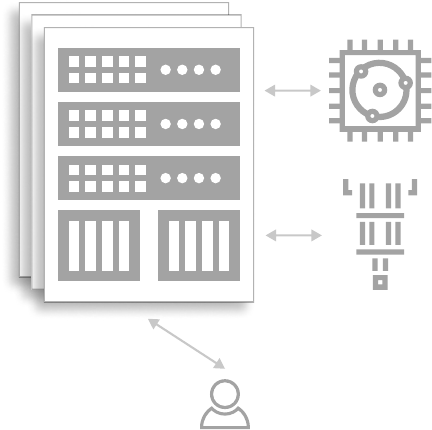}
        \caption{Co-located.}
        \label{subfig:hpcqc_integration_b}
    \end{subfigure}
    ~
    \begin{subfigure}[t]{0.3\textwidth}
        \centering
        \includegraphics[height=1.7in]{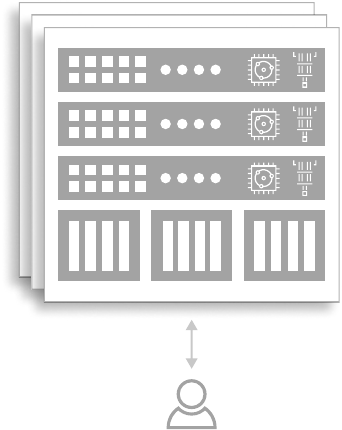}
        \caption{On-node.}
        \label{subfig:hpcqc_integration_c}
    \end{subfigure}
    \caption{All the possible integrations of QPUs in an HPC environment, a variation of the classification done by Elsharkawy et al.\ \cite{elsharkawy2023} that combines the two co-located models into one.}
    \label{fig:hpcqc_integration}
    \Description[Three possible schemes for QPUs integration]{Three possible schemes for QPUs integration, displaying visually the difference between having the QPUs completely separated from HPC, having them on-premises, but separated from the HPC classical resources, and having them completely integrated with the node.}
\end{figure*}

\begin{figure*}[t!]
    \centering
    \begin{subfigure}[t]{0.45\textwidth}
        \centering
        \includegraphics[height=2.7in]{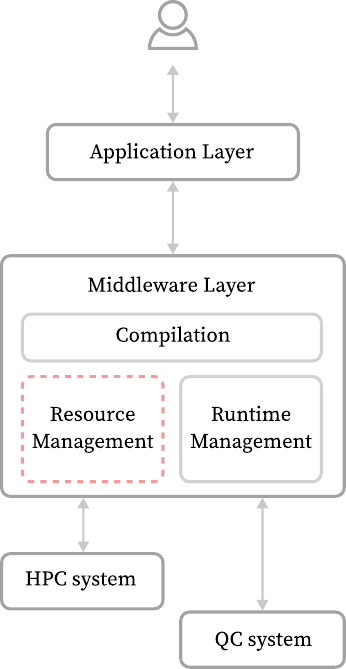}
        \caption{Resource management in the middleware~\cite{rallis2025hpcqcoverview}.}
        \label{subfig:quantumsw_a}
    \end{subfigure}%
    ~ 
    \begin{subfigure}[t]{0.45\textwidth}
        \centering
        \includegraphics[height=2.7in]{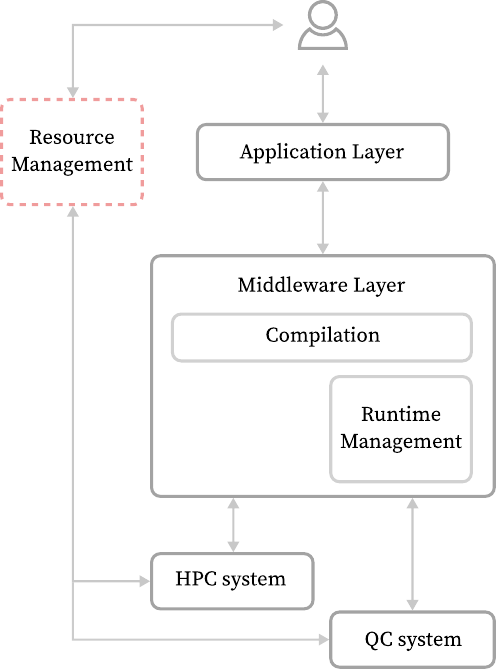}
        \caption{Resource management out of the middleware.}
        \label{subfig:quantumsw_b}
    \end{subfigure}
    \caption{Quantum software stack.}
    \label{fig:software_stack}
    \Description[Software stack scheme proposition]{Software stack scheme proposed based on the different approaches suggested in the bibliography. This scheme tries to generalize all of them and obtain the common parts in order to display the core needed for a software stack.}
\end{figure*}

With regard to quantum computing and HPC, in 2008 the work of Devitt et al.\ \cite{devitt2008high} marked the first attempt to consider QPUs within HPC environments. However, at that time, the goal was not to integrate quantum devices into such environments, but rather to envision the environments themselves as purely quantum. This model is called \emph{standalone} and can be visualized in Figure~\ref{subfig:hpcqc_integration_a}. The shift toward an integration perspective began with the work of Svore and Troyer~\cite{svore2016quantum} in 2016. They were the first ones envisioning QPUs as accelerators, a vision that has steadily gained traction over the years to the point where they are now treated outright as accelerators in many studies~\cite{Navaux2023}. Secondly, they anticipated cloud-based access to these devices, which would reside within HPC environments. This conceptualization of QPUs corresponds to the co-located model shown in Figure~\ref{subfig:hpcqc_integration_b}. Later that year, Humble and Britt~\cite{humble2016software} began to explore what such integration might look like and their article stands as one of the first ones to raise the central question of what the software stack for quantum computing in an HPC environment would be. In 2017, the theoretical exploration of how QPUs could fit into HPC centers continued~\cite{britt2017HPCQPU, britt2017quantum}, but it was not until 2021, with the work of Humble et al.~\cite{humble2021quantum}, that the idea of a software stack in a HPCQC context came to the forefront. A further step was taken in subsequent years with the works of Schulz et al.~\cite{schulz2022} and Seitz et al.~\cite{seitz2023Towards}, who began to examine how the preexisting classical software stack could be combined with the requirements imposed by quantum computing. Almost contemporaneously, the first well-defined example of such software stack appears~\cite{chichereau2023}, in which the GCC compilation stack was modified to include quantum features in the register allocation process. From 2023 to the present, both quantum computing and HPC have become rapidly expanding fields~\cite{garciabuendia2024survey} and, naturally, the combination of the two---usually referred to as HPCQC---also has, as shown in the work done by Döbler and Jattana~\cite{dobler2025survey}. During this period, the integration of QPUs into classical nodes in a GPU-like fashion began to be explored, giving rise to the on-node integration model---depicted in Figure~\ref{subfig:hpcqc_integration_c}. The reader is referred to two comprehensive reviews of the state of the art on the relationship between quantum computing and HPC for further details~\cite{pousa2024, rallis2025hpcqcoverview}. Following the one developed by Rallis et al.~\cite{rallis2025hpcqcoverview}, a summary of the software stacks presented so far can be seen in Figure~\ref{subfig:quantumsw_a}. In this scheme, resource management---both classical and quantum---is handled within the middleware. An alternative scheme, shown in Figure~\ref{subfig:quantumsw_b}, considers resource management as an user responsibility, thereby releasing the middleware from it. The relevance of this second scheme will become clear when CUNQA is introduced.

\begin{figure*}[t!]
    \centering
    \begin{subfigure}[t]{0.3\textwidth}
        \centering
        \includegraphics[width=0.9\textwidth]{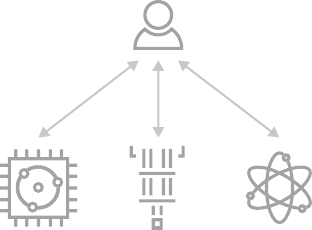}
        \caption{Embarrasingly parallel.}
        \label{subfig:dqc_embarrassing}
    \end{subfigure}%
    ~ 
    \begin{subfigure}[t]{0.3\textwidth}
        \centering
        \includegraphics[width=0.9\textwidth]{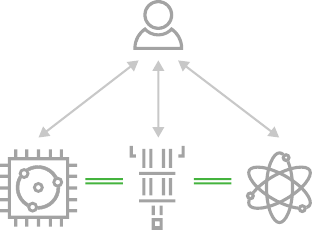}
        \caption{Classical communications.}
        \label{subfig:dqc_classical}
    \end{subfigure}
    ~
    \begin{subfigure}[t]{0.3\textwidth}
        \centering
        \includegraphics[width=0.9\textwidth]{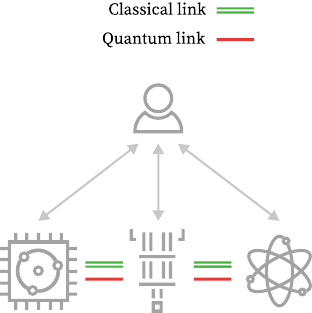}
        \caption{Quantum communications.}
        \label{subfig:dqc_quantum}
    \end{subfigure}
    \caption{All the possible DQC schemes.}
    \label{fig:dqc_schemes}
    \Description[Three possible schemes DQC]{Three possible schemes for DQC. The first one with no communications between QPUs, the second one with classical communications and the third one with quantum communications.}
\end{figure*}

When it comes DQC, a few explanations must be done in order to understand the context in which CUNQA is presented. Broadly speaking, two DQC schemes can be distinguished: distribution with communication and distribution without communication. The no-communication scheme, also referred to as embarrassingly parallel, can be seen in Figure~\ref{subfig:dqc_embarrassing} and corresponds to a classical distribution of quantum tasks, i.e., assigning different quantum tasks to different QPUs with the aim of parallelizing their execution. Hereafter, quantum tasks refers to any task of purely quantum nature together with its execution properties---such as the number of shots and, in the case of simulation, the corresponding simulation parameters. Regarding the communications scheme, there are two possible scenarios: having only classical connections between the QPUs, called classical-communication model, or having also quantum ones, called quantum-communication model. In the first case, shown in Figure~\ref{subfig:dqc_classical}, the same classical distribution of quantum tasks used in the no-communication model is applied, with the key difference that a QPU can receive classical information during execution. This introduces a synchronization problem---previously absent---, but enables the classical control of an instruction executed on a QPU based on results obtained from another QPU. In the second scenario, shown in Figure~\ref{subfig:dqc_quantum}, QPUs are connected through a quantum channel while maintaining the classical link from the previous model. Here, synchronization issues persist, together with additional challenges related to the physical layer, such as entanglement generation and distribution, the management of communication qubits dedicated to inter-QPU links and the underlying QPU connectivity. Despite these difficulties, this scheme enables larger computations by combining multiple QPUs, each with limited individual capability. This is also the strategy proposed in the aforementioned roadmaps to scale up the computational capacity of quantum systems. For a more detailed treatment of the subject, the reader is referred to two comprehensive reviews~\cite{barral2024, caleffi2024}.

It is under this whole scenario that CUNQA can be properly introduced. CUNQA is a DQC emulator for HPC, that is, CUNQA emulates DQC architectures within an HPC environment using its resources for this purpose. This approach involves several technical considerations, which are discussed in Section~\ref{sec:cunqa}. The discussion is complemented with use-case examples demonstrating the tool in operation, presented in Section~\ref{sec:display}. Section~\ref{sec:related_work} then compares CUNQA with existing tools to highlight its main contributions. Finally, Sections~\ref{sec:future_work} and~\ref{sec:conclusions} outline future directions for improvement and summarize the conclusions drawn from this work, respectively.

%% file: capitulos/2_cunqa.tex
\section{CUNQA}
\label{sec:cunqa}

The terms emulation and simulation are often used interchangeably, yet they are not the same. Emulation seeks to faithfully replicate a system, whereas simulation aims to reproduce its behavior. In classical computing, emulating a device typically involves translating the original program’s instructions from the emulated device into native ones understood by the host machine. If a quantum computer was to be emulated by a classical computer, no quantum instructions could be translated to classical ones because the computation paradigms are completely different. The natural approach is to follow what Willisch~\cite{willisch2020thesis} proposed: the classical counterpart should solve the physical equations governing the quantum system in order to replicate faithfully its behavior, therefore, emulating it. 

Describing CUNQA as a DQC emulator for HPC may suggest that the emulation focuses on the QPUs themselves, but this is not the case. What CUNQA aims to emulate are the three DQC models shown in Figure~\ref{fig:dqc_schemes}, not the QPUs individually. Accordingly, an extended definition would be: CUNQA is a DQC emulator for HPC with simulated QPUs. The ``for HPC'' part also implies some considerations as shows that CUNQA was conceived from the outset within an HPC context. Specifically, CUNQA supports the emulation of the co-located scheme, shown in Figure~\ref{subfig:hpcqc_integration_b}, and the on-node scheme, shown in Figure~\ref{subfig:hpcqc_integration_c}, while discarding the standalone model of Figure~\ref{subfig:hpcqc_integration_a} as obsolete and incompatible with the accelerator paradigm.

CUNQA is the first tool to emulate the three DQC schemes in an HPC environment following the co-located and on-node models. To clarify how this is accomplished, each layer of the software stack is described in this section. It has to be noted that the software stack structure follows the scheme shown in Figure~\ref{subfig:quantumsw_b}, where resource management is the user’s responsibility. The lowest layer of the stack, presented in Section~\ref{subsec:virtual_qpus}, corresponds to the simulated QPUs—hereafter referred to as virtual QPUs (vQPUs). Section~\ref{subsec:resource_management} then describes resource management, focusing primarily on how vQPUs are handled. Once the vQPUs and their management are established, Section~\ref{subsec:cunqa_api} introduces the interface through which users interact with them and, therefore, describing the application layer. Finally, Section~\ref{subsec:middleware} explains how the interface and the vQPUs are connected by explaining the middleware layer. 

From a programming perspective, CUNQA is implemented in both Python and C++, and each of the aforementioned sections specifies which language is used in each case.

\subsection{System layer: Virtual QPUs (vQPUs)}
\label{subsec:virtual_qpus}
\begin{figure*}[t!]
    \centering
    \begin{subfigure}[t]{0.47\textwidth}
        \centering
        \includegraphics[width=0.95\linewidth]{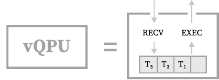}
        \vspace{0.2cm}
        \caption{vQPUs for no- and classical-communication models.}
        \label{subfig:virtual_qpu}
        \Description[Insides of a vQPU]{The image shows the scheme of a vQPU. It displays the two threads described before: one that receives and one that executes the information, along with the queue in which the former one stores the receive quantum task for the latter to execute it and return the results.}
    \end{subfigure}
    \hfill
    \begin{subfigure}[t]{0.47\textwidth}
    \centering
        \includegraphics[width=0.95\linewidth]{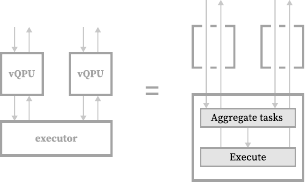}
        \vspace{0.2cm}
        \caption{vQPUs for quantum-communication model.}
        \label{subfig:qpus_executor}
        \Description[Quantum communication executor.]{The image shows how the so called ``executor'' processes manages and executes the stream of instructions that receives from the vQPUs.}
    \end{subfigure}
    \caption{In \ref{subfig:virtual_qpu} the vQPUs are responsible for the execution while in \ref{subfig:qpus_executor} the vQPUs delegate in an executor process.}
    \label{fig:vqpus_schemes}
\end{figure*}

A vQPU is a classical process running on a HPC environment with an allocated set of classical resources responsible of simulating the behavior of a real QPU. As a result, it is entirely independent of the quantum tasks executed on it. Users can interact with it as if it was a real QPU available in the system: they can query its status, check its availability and submit quantum tasks for execution. In essence, it behaves as an actual accelerator, that is, a device that communicates with the CPU to receive tasks, execute them and return the results. This concept is not entirely new, as instances of its use can be found in the literature~\cite{claudino2024}. In this work, a vQPU comprises two main components that can be replaced or extended, making them decoupled of any specific third-party implementation:

\begin{itemize}
    \item \emph{Server component}: responsible for the communications of the vQPUs with the user. This component handles the reception of quantum tasks, their storage in a queue and the delivery of their results back to the user. Given their classical client-server structure, a standard socket library is employed---in the current version of CUNQA these can be Asio or ZMQ.
    
    \item \emph{Simulator component}: performs the actual execution of quantum tasks. At present, three simulators are supported: AerSimulator 0.16.0~\cite{aer}, a specific commit of MQT-DDSIM~\cite{munich} and CunqaSimulator---a simple own simulator developed for testing purposes~\cite{cunqasimulator}. In the no-communication case, the simulator is left to do all the simulation work, being a black box from CUNQA perspective. However, when there is communications between vQPUs, either classical or quantum, the way the simulation is performed had to be modified. In such cases, the simulator process iterates over all shots, each shot applying the circuit instructions extracted from the simulator one after the other and finally all shots' results are manually assembled to return the post-processed result to the user. This method allows maximum control for adding the new custom communication instructions at low level. This extraction of gates outside the simulator canonical path causes the additional complexity of simulating the noise, so, the version of CUNQA presented in this work only supports noise in the no-communication model.
\end{itemize}
Each vQPU operates through two threads: one in charge on server component and and another in charge on the simulation. This vQPU architecture and execution flow is illustrated in Figure~\ref{subfig:virtual_qpu} and they remain valid across the no-communication and classical-communication models presented in Figure~\ref{fig:dqc_schemes}. When a vQPU supports classical communications, the simulator component has an additional element: the classical communications channel. This channel, available with ZMQ and MPI, is the responsible of sharing classical bits between vQPUs at runtime using a classical send-receive scheme.

A difference arises for the quantum-communication case: quantum links may generate entanglement between qubits located in different vQPUs. Then, quantum tasks that involve quantum communications cannot be simulated independently on each vQPU: they have to be simulated within a unique process. Figure \ref{subfig:qpus_executor} shows what does this mean at low level: vQPUs does not have a proper simulator component anymore, but they are now just responsible of send their quantum task part to a new \textit{executor} process and receive back the result. The simulation component falls on the executor which is also responsible for receiving and correctly coordinating their joint execution. The result is returned to the user as if the circuit had been executed on a single vQPU---that is, with aggregated counts. This approach reflects how execution would take place in a quantum multicore architecture.

\begin{figure*}[t]
    \centering
    \begin{subfigure}[b]{0.45\linewidth}
        \centering
        \begin{adjustbox}{scale=0.9}
        \begin{quantikz}[wire types={n,n,n,n}]
             &\lstick{$|a\rangle$}           &  \setwiretype{q} & \ctrl{1}  & \gate{H} & \meter{}         & \cwbend{2} \setwiretype{c} \\
             \gategroup[wires=2, steps=9, style={dashed, rounded corners, fill=gray!10, inner sep=2pt}, background]{} 
             &                               & \lstick[2]{$|\Phi^+\rangle$}     & \targ{} \setwiretype{q}       & \meter{}     \\
             &                               &                  & \setwiretype{q}        & \gate{X} \vcw{-1} & & \gate{Z}   & \swap{1} &      \\
             &\lstick{$|0\rangle$}           & \setwiretype{q}  &       &                &         & &\targX{} &     
        \end{quantikz}
        \end{adjustbox}
        \caption{Teledata.}
        \label{subfig:teledata}
    \end{subfigure}
    \begin{subfigure}[b]{0.45\linewidth}
        \centering
         \begin{adjustbox}{scale=0.9}
         \begin{quantikz}[wire types={n,n,n,n}]
             &\lstick{$|a\rangle$}           &  \setwiretype{q}                 & \ctrl{1}                  &                   &           &           & \gate{Z} \vcw{2} &    \\
             \gategroup[wires=2, steps=9, style={dashed, rounded corners, fill=gray!10, inner sep=2pt}, background]{} 
             &                               & \lstick[2]{$|\Phi^+\rangle$}     & \targ{} \setwiretype{q}   & \meter{}                                                      \\
             &                               &                                  & \setwiretype{q}           & \gate{X} \vcw{-1} & \ctrl{1}  & \gate{H}  & \meter{}  &       \\
             &\lstick{$|t_1\rangle$}         & \setwiretype{q}                  &                           &                   & \control{}  &           &   &     
        \end{quantikz}
        \end{adjustbox}
        \caption{Telegate.}
         \label{subfig:telegate}
    \end{subfigure}
    \caption{Quantum communication protocols extracted from the work of Barral et al.\ \cite{barral2024} with the communication qubits highlighted.}
    \label{fig:circuits_teledata_telegate}
    \Description[Quantum communication protocols]{In this figure, both teledata and telegate protocols are depicted with the two communication qubits highlighted.}
\end{figure*}
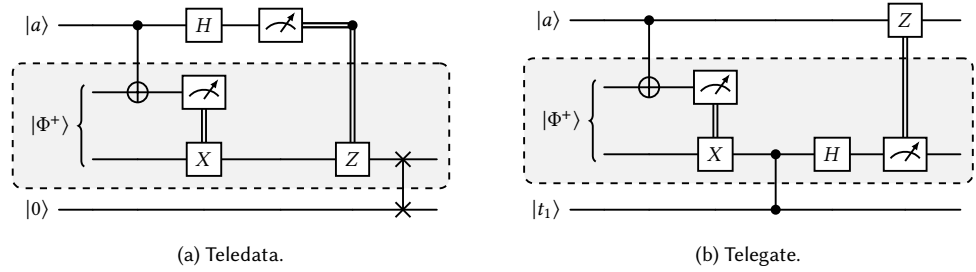

The main difference between this executor process with respect to the execution of an usual vQPU, is that its associated simulator will execute a single circuit with as many qubits as the sum of the qubits of all the circuits sent to be executed, plus two qubits reserved for the communication protocols. These two qubits come from the two communication protocols implemented: teledata and telegate. Figure~\ref{fig:circuits_teledata_telegate} shows how these protocols employ two qubits to generate an EPR pair shared between the two QPUs. This way, each time a protocol is executed, an EPR pair is created using the same two qubits. Recycling communication qubits allows for indefinite quantum communications without having to increase the number of qubits to be simulated.

\subsection{Resource management}
\label{subsec:resource_management}
As vQPUs are classical resources with the goal of simulating a real QPU, CUNQA implements a set of commands that handle their life cycle---launching, checking availability, terminating and releasing their associated resources. The classical nature of a vQPU within an HPC environment implies that these commands are implemented as SLURM wrappers with a set of additional options. This is coherent with the software stack structure selected for CUNQA at the beginning of the section: the resource management is a user responsibility as the user must first reserve the necessary resources by launching the vQPUs. In other words, by the time the user executes the program defined at the application layer, both the classical and quantum resources---represented here by the vQPUs---must already be managed. 

For launching vQPUs, CUNQA provides the \texttt{qraise} bash command, which allows fine-tuning of the vQPUs’ characteristics through several configurable flags. A generic example of the \texttt{qraise} command is shown below:
$$ \texttt{qraise -n 4 -t 00:10:00 (MORE OPTIONAL FLAGS)} $$

The two flags shown are mandatory and specify the number of vQPUs to launch (\texttt{-n}) and the maximum time they remain available (\texttt{-t}). Each exeution of the \texttt{qraise} command deploys vQPUs in a group referred to as \emph{family}. Each family defines a homogeneous set of vQPUs that share the same configuration. The optional flags help the user tune additional vQPU properties, classical resources configuration, DQC model and the integration with the HPC environment.

\begin{itemize}

    \item \emph{Simulated QPU characteristics}. In order to model the behavior of a real QPU, the number of qubits, their coupling graph, supported quantum instructions and many other traits are needed. All those are commonly grouped and referred to as \emph{backend}, which will be provided via the \texttt{-{}-backend} flag, taking the path to a JSON file describing it. A key feature of a QPU backend is the set of noise properties of its qubits, which can be differently specified by the \texttt{-{}-noise-prop} option, which also expects the path to a JSON file. From these noise properties, CUNQA builds the noise model according to the simulator instructions via Qiskit, justified by the fact that the purpose of CUNQA is to emulate DQC and not focus on simulating vQPUs. By default, the backend is noiseless with 32 qubits. For a more detailed example of the JSON describing the backend and for additional noise-related flags, the reader is referred to the CUNQA documentation~\cite{cunqa_docs}.
    
    \item \emph{Simulator}. Using the \texttt{-{}-sim} flag, the user can specify the simulator to be used by the vQPU. The default selection is AerSimulator, which for the current version is the only one that supports noise models.

    \item \emph{Classical resources}. Flags such as \texttt{-c}, \texttt{-{}-mem-per-qpu} and \texttt{--n\_nodes} are supported. These are related to the resource management options provided by the standard SLURM commands.
    
    \item \emph{DQC model}. Flags \texttt{-{}-classical\_comm} and \texttt{-{}-quantum\_comm} are used to enable these communications among the raised vQPUs, this way users are allowed to build the classical- and quantum-communication models, respectively. If none of them is specified, the no-communication scheme is deployed.
    
    \item \emph{QC and HPC integration scheme}. This option determines whether the user can access the deployed vQPUs only if located on the same node or whether it can access to them independently the located node, i.e., on-node and co-located schemes. Default option is the first one and the latter is activated through the \texttt{-{}-co-located} flag.
    
    \item \emph{Name}. A name can be assigned to a family of vQPUs for its identification.
\end{itemize}

Other implemented commands include \texttt{qdrop} and \texttt{qinfo}. The former, a wrapper of SLURM’s \texttt{scancel} command, frees the resources allocated to vQPUs. The latter provides information about the vQPUs currently active in the system. 

Although it is not intended to be the main workflow in CUNQA, all these bash commands are also available through the user interface that, as it is presented in the next section, is a Python API. This allows users to launch and terminate vQPUs before and after executing specific tasks. This enables both execution and resource management to be handled at once, while still sticking with the model described in Figure~\ref{subfig:quantumsw_b}.

\subsection{Application layer: \texttt{cunqa} package} 
\label{subsec:cunqa_api}
To interact with the vQPUs, CUNQA provides a Python API to design, submit to vQPUs and receive the results of quantum tasks. These functionalities can be summarized in two main concepts: design and execution. The design is handled by the \texttt{CunqaCircuit} Python class, which provides the necessary tools to model quantum tasks within CUNQA’s capabilities. The execution, in turn, relies on a set of methods and classes, each with a clearly defined role within the execution chain.

\subsubsection{CunqaCircuit} Starting with the circuit creation capabilities, CUNQA provides its own circuit class, \texttt{CunqaCircuit}, bearing a strong resemblance in syntax to Qiskit's \texttt{QuantumCircuit}~\cite{qiskit2024}. Together with QASM2, they are the three circuit formats supported by CUNQA. \texttt{CunqaCircuit}'s main purpose is to implement the distributed instructions, in line with the DQC framework that CUNQA operates under. Below there is an example of a \texttt{CunqaCircuit} instance
\begin{lstlisting}[style=onelinepython]
    circuit = CunqaCircuit(num_qubits = 3, num_clbits = 3, id = "cunqacircuit")
\end{lstlisting}
where \texttt{num\_qubits} is the number of quantum registers of the circuit, \texttt{num\_clbits} is the number of classical registers and \texttt{id} is a circuit identifier, the basic tool to create distributed instructions. 

To illustrate how to communicate circuits between each other, hereafter, \texttt{send\_circuit} and \texttt{recv\_circuit} are two different instances of \texttt{CunqaCircuit} having ``send\_circuit'' and ``recv\_circuit'' as identifiers, respectively.

\begin{itemize}
    \item \textbf{Classical communications.} To send the measurement result from a circuit to another, the method 
\begin{lstlisting}[style=onelinepython]
    send_circuit.measure_and_send(control_qubit, target_circuit)
\end{lstlisting}
    must be applied to the sending circuit, where the \texttt{control\_qubit} argument means the qubit index where the measurement will be performed and \texttt{target\_circuit} the identifier of the receiving circuit. This circuit must be waiting for the classical bit, and to do this the method 
\begin{lstlisting}[style=onelinepython]
    recv_circuit.remote_c_if(gate, target_qubits, control_circuit)
\end{lstlisting}
    must be applied by it. In this case, the argument \texttt{gate} is the name of the gate to be applied if the receiving bit is equal to one, \texttt{target\_qubits} are the index of the qubits where the gate will be applied, \texttt{control\_circuit} is the identifier of the sending circuit and \texttt{params} are the parameters in case the applied gate is parametric.
    
    \item \textbf{Quantum communications: teledata.} In this protocol, the qubit's state in the sending circuit is transferred to the state of a qubit on the receiving circuit. The state is destroyed on the sending circuit, so no copy of states occurs, in agreement with the no-cloning theorem. To implement the teledata, the method 
\begin{lstlisting}[style=onelinepython]
    send_circuit.qsend(send_qubit, target_circuit)
\end{lstlisting}
    must be applied on the sending circuit, specifying with the \texttt{send\_qubit} argument the index of the qubit in which the state will be transmitted and \texttt{target\_circuit} as the identifier of the receiving circuit. On this receiving side, the method 
\begin{lstlisting}[style=onelinepython]
    recv_circuit.qrecv(recv_qubit, control_circuit)
\end{lstlisting}
    must be applied with the \texttt{recv\_qubit} argument as the index of the receiving qubit and the \texttt{control\_circuit} identifier of the circuit that is sending the qubit.
    
    \item \textbf{Quantum communications: telegate.} In this protocol, a qubit from the sending circuit remotely controls gates in a receiving circuit without transferring its state completely, which allows the recovery of the initial state of the control qubit. Therefore, this protocol can be thought of in three blocks: the first, known as \textit{cat-entanglement}, transfers the necessary information from one circuit to another; the second consists of applying the controlled gates to the receiving circuit and, the third, known as \textit{cat-disentanglement}, in which the control qubit recovers its initial state, see~\cite{barral2024}. 
    Thus, the implementation that best fits this protocol is a Python \textit{context manager}, which applies the cat-entangler upon entering and the cat-disentangler upon exit. Explicitly, 
\begin{lstlisting}[style=onelinepython]
    with send_circuit.expose(control_qubit, target_circuit) as rcontrol:
        recv_circuit.controlled_gate(rcontrol, target_qubit)
\end{lstlisting}
    where the qubits indexes, the controlled gates and circuit identifier of the receiving circuit must be specified.
\end{itemize}

Additionally, \texttt{CunqaCircuit} implements convenient operators to build large circuits out of smaller ones. The sum of two \texttt{CunqaCircuit} objects, written as \texttt{circ\textsubscript{0}+circ\textsubscript{1}}, returns a deeper circuit with the instructions of the first summand followed by those of the second. The union of two \texttt{CunqaCircuit} objects, written as \texttt{circ\textsubscript{0}|circ\textsubscript{1}}, returns a new circuit with its first quantum registers with their gates being those of the \texttt{circ\textsubscript{0}} and and similarly the last quantum registers with their gates being those of the \texttt{circ\textsubscript{1}}. On the other direction, it is possible to extract pieces of a given circuit using the methods \texttt{hor\_split} and \texttt{vert\_split}, which divide a circuit with horizontally or vertically cuts, after a given qubit or after a given position respectively. Other methods like \texttt{len}, which returns the circuit depth, or \texttt{contains}, which returns a boolean specifying if a gate is present in the circuit, are available, while more functionalities are constantly being added. See CUNQA documentation~\cite{cunqa_docs} for further details. 

\subsubsection{Job submission and results reception.} To run the created quantum circuits on a vQPUs, firstly is needed to access to the deployed QPUs. This is obtained through the \texttt{get\_QPUs} function, which returns a list of instances of the \texttt{QPU} class with each one containing a client connected to the server of a vQPU. This function allows two arguments: one for disabling the \emph{on-node} through the boolean argument \texttt{on\_node} and the other to filter the vQPUs through its family name, both introduced in Section~\ref{subsec:resource_management}. 

To run a circuit on the no-communication model, the user chooses one of the \texttt{QPU} instances and calls its method \texttt{QPU.run(circuit, options)}, where the \texttt{options} argument includes number of shots, simulation method, circuit compilation settings, etc. This call sends the information in an asynchronous way, making possible the parallelization of the executions when using more than one vQPU. 

On the other hand, to run a circuit in a model with communications---either classical or quantum---the function \texttt{run\_distributed(circuits, qpus, options)} must be used, where \texttt{circuits} and \texttt{qpus} are lists of quantum circuits and \texttt{QPU} instances, respectively. The number of \texttt{QPU} instances must be greater or equal than the number of circuits to submit. Additionally, the associated vQPUs to the \texttt{QPU} instances must have been deployed with the desired communications on. 

Both methods, \texttt{run} and \texttt{run\_distributed}, return instances of the class \texttt{QJob} which handles the interaction with the asynchronous simulation taking place in the vQPUs. The results of the circuit execution are attained using the attribute \texttt{QJob.result}, a Python dictionary including the counts of the measures (\texttt{counts}) and the time taken to perform the simulation (\texttt{time\_taken}). The \texttt{result} attribute provokes a blocking call where the system must wait for the execution to finish before continuing to the next line. For convenient extraction of results of multiple simulation jobs the function \texttt{gather} was created, taking as inputs a list of \texttt{QJob} instances and returning a list of \texttt{Result} objects.

Finally, it must be noted that the \texttt{QJob} class possesses the \texttt{upgrade\_parameters}. Some algorithms---such as Variational Quantum Algorithms (VQAs)---include a quantum part consisting of a parametric circuit executed iteratively with a different set of parameters each time. In such cases, each new circuit differs from the previous one only in the values of certain parameters, not in its overall structure. The \texttt{upgrade\_parameters} method enables the update of these values, thereby avoiding the transmission of redundant information in each iteration of the algorithm.

\subsection{Middleware}
\label{subsec:middleware}

The middleware is commonly understood as the ``software glue'' that connects two software layers, but this definition lacks precision. In this work, \emph{middleware} is meant to be the set of software tools dedicated to connect the Python interface with the vQPUs, written in C++. In CUNQA, this connection is carried out mainly by two procedures: transforming the quantum circuits into an intermediate representation (IR) and connecting the user with the vQPUs. On the one hand, CUNQA Python package converts the supported circuit representations---described in Section~\ref{subsec:cunqa_api}---into a JSON format with a predefined scheme. JSON was the chosen IR of quantum circuits motivated by the fact that simulators such as AerSimulator already rely on such a representation. On the other hand, the communications between the user and the vQPUs are implemented in C++. The user side is also compiled for its use in Python, combining C++ performance with Python-level accessibility through the API. The connection between this interface and the lower level is made through a single point: the vQPU client presented above in Section~\ref{subsec:virtual_qpus}. This unique contact point has the great advantage of decoupling both Python and C++ layers. An user can use CUNQA through the Python interface without worrying about the low-level part.

It may be inferred from the previous description of the middleware that the compilation stage---which usually comprises several optimization steps of the circuit---is, simply, transforming the circuit representations to JSON format with a predefined scheme. But this is not the case, as CUNQA supports Qiskit transpilation of circuits, leveraging its circuit optimization as part of the API. No optimization method, nor any advance compilation tool is developed as a core part of CUNQA. This is due to the fact that CUNQA's purpose is not the optimal compilation of quantum circuits, but the integration of DQC into HPC environments. Overall, the middleware’s simplicity is a deliberate design decision intended to concentrate development efforts on the emulation process and on the orchestration of the DQC architectures implemented in CUNQA.

%% file: capitulos/3_display.tex
\section{CUNQA on display: the QPE algorithm}
\label{sec:display}

To prove that CUNQA can emulate the three DQC schemes introduced, it was decided to run the same algorithm adapted to each of the three schemes. The Quantum Phase Estimation algorithm (QPE) was selected for this purpose. The selection of only one algorithm is based on the idea of being able to faithfully compare the models, avoiding the external differences brought by the selection of different algorithms. 

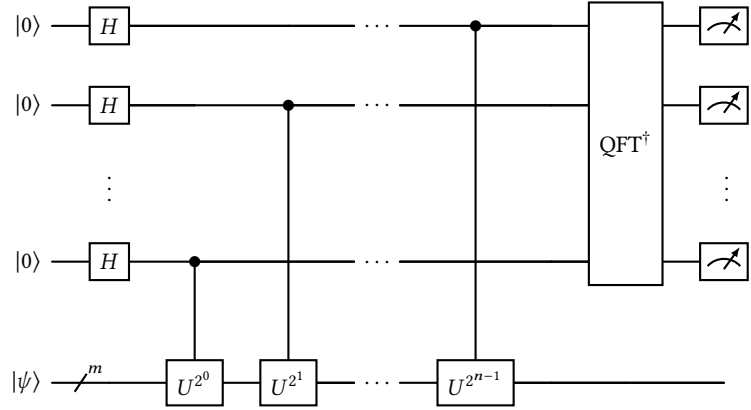
\begin{figure}[!t]
    \centering
    \begin{quantikz}[wire types={q,q,n,q,q}]
        \lstick{$\ket{0}$}    & \gate{H}     & \qw            & \qw            & \qw \text{ } \ldots \text{ }  & \ctrl{4} & \qw     & \gate[4]{\text{QFT}^{\dagger}} & \meter{} \\
        \lstick{$\ket{0}$}    & \gate{H}     & \qw            & \ctrl{3}       & \qw \text{ } \ldots \text{ }  & \qw                & \qw                            & \qw      & \meter{} \\
                              & \ \vdots\    &                &                &                               &                    &                                &          & \ \vdots \\
        \lstick{$\ket{0}$}    & \gate{H}     & \ctrl{1}       & \qw            & \qw \text{ }  \ldots \text{ } & \qw                & \qw                            & \qw      & \meter{}\\[0.5cm]
        \lstick{$\ket{\psi}$} & \qwbundle{m} & \gate{U^{2^0}} & \gate{U^{2^1}} & \qw \text{ }  \ldots \text{ } & \gate{U^{2^{n-1}}} & \qw                            & \qw      &  \qw 
    \end{quantikz}
    \caption{QPE circuit}
    \label{fig:QPE}
    \Description[QPE algorithm]{QPE circuit displayed.}
\end{figure}

The QPE---first introduced in 1995 by Kitaev~\cite{QPE}---aims to estimate eigenvalues from unitary operators. Specifically, the algorithm will approximate the phase $\phi \in [0,1)$ of the eigenvalue $e^{2\pi i \phi}$ of a unitary operator $U$. Hereafter, $U$ will be an arbitrary unitary operator and $|\psi\rangle$ one of its eigenvectors such that $U |\psi\rangle = e^{2\pi i \phi} |\psi\rangle$. The quantum circuit---displayed in Figure~\ref{fig:QPE}---is composed by two quantum registers: one for the $n$ ancilla qubits that will store the phase information and another where the controlled $U^{2^t}$ operators are applied, for $t\in \{0, ..., n-1\}$. The phase $\phi$ is the estimated as 
$$\phi\simeq \hat{\phi} = \frac{\xi}{2^n} \text{ with an error of } \frac{1}{2^{n}},$$
where $\xi$ is the decimal representation of the most frequent outcome bit string of the former quantum register.

For obtaining a result with high precision, one can increase the number $n$ of ancilla qubits increasing the number of possible outcomes up to $2^n$, which forces to increase the number of shots required by the algorithm. This conforms a clear case were de no-communication model of CUNQA can be applied: distribution of shots among the vQPUs. The classical communication model fits on the so called Iterative Quantum Phase Estimation Algorithm (IPEA), that leverages classical communications between QPUs to reduce the number of ancilla qubits. Finally, the fact that the QPE is separated in two quantum registers allows to easily separate the QPE into two sets with a well known number of controlled gates affected. This fits in the quantum-communication model as the controlled gates required to be distributed are perfectly delimited. 

The rest of this section will show how the QPE is implemented under the aforementioned three models with a final discussion of the results. The selected unitary operator $U$ is the $R_z$ operator described as 
$$ R_z(2\theta) = \begin{pmatrix} e^{-i\theta} & 0  \\ 0 &  e^{i\theta} \\ \end{pmatrix}, $$
and its selected eigenvector is $|\psi\rangle = |1\rangle$. The $R_z$ was chosen because the simple preparation of its eigenvectors $\{|0\rangle, |1\rangle\}$, one of the major bottlenecks of QPE. The chosen eigenvector is $|1\rangle$ because it allows to know the result in advance through the simple relationship $\theta = 2\pi\phi$. 

\subsection{No-communication: distribution of shots}
\label{subsec:no-comm}

The no-communication case, as previously noted, is the most straightforward one: classical distribution is applied to quantum tasks. In this case, the distribution consists of dividing the total number of shots by the number of available vQPUs and assigning each vQPU its corresponding share. This naive technique is referred to as distribution of shots. By applying it, a quantum task that would otherwise execute all shots sequentially can now parallelize execution in proportion to the number of vQPUs available. Figure \ref{code:QPE} presents the Python code that implements this case, employing all the different methods presented in Section~\ref{subsec:cunqa_api}. From lines 1-12, the circuit is constructed; after this, in line 15, \texttt{get\_QPUs} is employed to get the vQPUs available as a list of \texttt{QPU} objects; finally, in line 19, the asynchronous \texttt{run} method is called iteratively to send the quantum tasks to the corresponding vQPUs.

\begin{figure}[!t]
    \centering
\begin{lstlisting}[style=mypython]
    # Defining the QPE circuit
    qpe = CunqaCircuit(num_qubits + 1, num_qubits)
    qpe.x(num_qubits)

    for i in range(num_qubits):
        qpe.h(i)                    
        for _ in range(2**i):
             qpe.crz(phase, i, num_qubits) 
             
    # Inverse QFT on first n-1 qubits (not available in the API, abbreviated for clarity)
    qpe.IQFT(0, num_qubits - 2)
    qpe.measure_all()

    # Obtaining available QPUs and distributing the shots
    qpus = get_QPUs()
    shots_per_qpu = [shots / size(qpus) for i in range(size(qpus))] 

    # Submitting the jobs and gathering results
    qjobs = [qpu.run(qpe, shots_per_qpu) for qpu, qpu_shots in zip(qpus, shots_per_qpu)]
    results = gather(qjobs)
    
\end{lstlisting}
    \caption{Code for the QPE with distributed \textit{shots}.}
    \label{code:QPE}
    \Description[QPE execution in CUNQA without communications]{Python file designing the QPE circuit and executing it using CUNQA.}
\end{figure}

\subsection{Classical-communication: Iterative Phase Estimation Algorithm (IPEA)}
\label{subsec:classical-comm}

\begin{figure}[!t]
    \centering
    \adjustbox{scale=0.75}{
        \begin{quantikz}[wire types = {n, n, q, q}]
             & & & &\slice[style = blue]{} & & & & &\lstick{$\ket{0}${ }}\slice[style = blue]{} & \gate{H} \setwiretype{q}& \ctrl{3} &\gate{R_z(-\frac{\pi}{2^2})} &\gate{R_z(-\frac{\pi}{2})} &\gate{H} & \meter{} \\
             & & & &\lstick{$\ket{0}${ }} & \gate{H} \setwiretype{q}& \ctrl{2} &\gate{R_z(-\frac{\pi}{2})} & \gate{H} & \meter{} & \setwiretype{c} & & & \ctrl{0}\wire[u][1]{c}\\
            \lstick{$\ket{0}${ }} & \gate{H} & \ctrl{1}  & \gate{H} &\meter{} & \setwiretype{c} & &  \ctrl{0}\wire[u][1]{c}\ & & & & &  \ctrl{0}\wire[u][2]{c}\\
            \lstick{$\ket{\psi}$} &    & \gate{U^{2^2}} &   &   &  & \gate{U^{2^1}} &   &   & & & \gate{U^{2^0}} & 
        \end{quantikz}
    }
    \caption{QPE with measures before the QFT controls, revealing the IPEA structure.}
    \label{fig:IPEA}
    \Description[IPEA circuit]{Reorder of the QPE algorithm revealing the IPEA structure. Each block is of the algorithm is delimited.}
\end{figure}
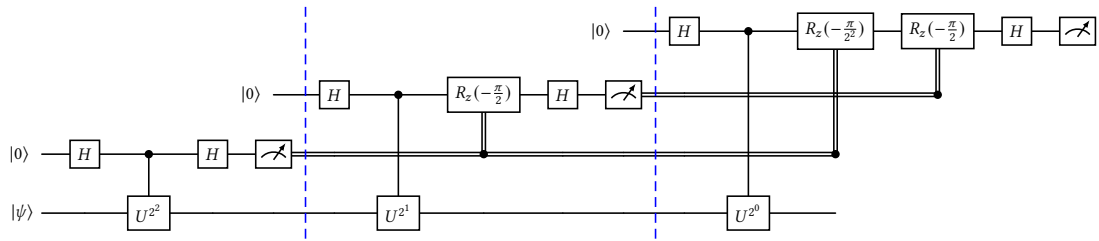

The Iterative Phase Estimation Algorithm (IPEA)~\cite{ipea_article} addresses the problem of increasing the number of qubits for gaining precision in the QPE. IPEA reduces the amount of ancilla qubits by sequentially computing each bit of the binary expansion of the phase. The size of the circuits employed is of $(1 + m)$ qubits versus the $n + m$ qubits required in the QPE---where $2^{m}$ is the dimension of the unitary matrix $U$. This reduction is achieved through the introduction of classical communications, which allow a circuit to transmit its computed bit to all successive circuits, keeping all information available at each step.

To obtain the IPEA structure from the QPE, first notice that the last operation on the ancilla qubits before measurement is the inverse QFT block, which consisted on a ladder of Hadamards and controlled rotations. By the \emph{Deferred Measurement Principle}~\cite{deferred_meas},~\cite[\S 4.4]{nielsen2000} this is equivalent to measuring before the controlled rotations, resulting on the circuit shown in Figure~\ref{fig:IPEA} with classically  controlled gates. These classical controls suggest a circuit partition in several vQPUs with classical communications. This partition is shown in Figure~\ref{fig:IPEA} with blue dotted lines, determining the distributed IPEA that this section focuses on. Notice that the partition is possible as in the controlled unitary register each gate returns the eigenvector making each step independent of the others --- if the eigenvector can be reliably prepared.

\begin{figure}[!t]
    \centering
\begin{lstlisting}[style=mypython]
    # Getting the available vQPUs
    qpus_QPE  = get_QPUs()
    
    # Defining the IPEA circuits
    circuits = []
    for i in range(num_qpus): 
        theta = 2 * pi * 2**(num_qpus - i - 1) * phase # pi = 3,141592...
    
        circuits[i] = CunqaCircuit(qubits=2, cbits=2, id= f"circuit_{i}")
        circuits[i].h(0)
        circuits[i].x(1)
        circuits[i].crz(theta, 0, 1) # Powers of the controlled unitary

        # Receiving bits from previous circuits
        for j in range(i):
            qpe_angle = -pi * 2**(j - i)
            circuits[i].remote_c_if("rz", qubits = 0, 
                                    param = qpe_angle, control_circuit = f"circuit_{j}")

        circuits[i].h(0)

        # Sending bits to upcoming circuits
        for k in range(num_qpus - i - 1):
            circuits[i].measure_and_send(qubit = 0, target_circuit = f"circuit_{i + k + 1}") 
    
        circuits[i].measure_all()
    
    # Submitting the jobs and gathering results
    distributed_jobs = run_distributed(circuits, qpus_QPE)
    results = gather(distributed_jobs)
\end{lstlisting}
    \caption{IPEA implementated on CUNQA}
    \label{code:IPEA}
    \Description[IPEA python code]{Python file designing the IPEA circuit and executing it using CUNQA.}
\end{figure}

\begin{figure*}[t!]
    \centering
    \begin{subfigure}[b]{0.47\textwidth}
        \centering
        \begin{quantikz}[wire types={n, n,q,q}]
            & & & \vdots & & \\[-0.35cm]
            & & & & & \\
            \lstick{$\ket{0}${ }}   & \gate{H} & \ctrl{1}       &  \gate{R_z}\wire[u][1]{c}{\tilde{b}_{0}, \tilde{b}_{1}, \dots, \tilde{b}_{i-1} \hspace{0.1cm}} & \gate{H} &\meter{}\wire[d][1]{c}{\tilde{b}_i}\\
            \lstick{$\ket{\psi}$}  &      & \gate{U^{2^{n-i-1}}} &               &   & \setwiretype{n}\vdots \\
        \end{quantikz}
        \caption{Circuit structure for the $i$-th step of the IPEA.}
        \label{subfig:IPEA_block}
    \end{subfigure}%
    ~ 
    \begin{subfigure}[b]{0.47\textwidth}
        \centering
        \includegraphics[width=0.70\linewidth]{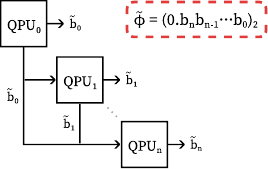}
        \caption{Bit flow between vQPUs in the IPEA.}
        \label{subfig:IPEA_qpus}
    \end{subfigure}
    \caption{The bit $b_{i}$ in the binary expansion of $\tilde{\phi}$ is the most frequent bit $\tilde{b}_{i}$ after the shots.}
    \Description[Three possible schemes DQC]{Three possible schemes for DQC. The first one with no communications between QPUs, the second one with classical communications and the third one with quantum communications.}
\end{figure*}

Figure \ref{code:IPEA} shows Python code implementing the IPEA using CUNQA. It retrieves the available QPUs with \texttt{get\_QPUs} in line 2, creates as many IPEA circuits as QPUs are deployed in lines 5 to 25, runs them using \texttt{run\_ditributed} in line 28 and collects the results using \texttt{gather} in line 29.
The circuit creation logic is abstracted away in the generic Figure \ref{subfig:IPEA_block}: a Hadamard gate is applied to the unique ancilla qubit to enter superposition, the controlled unitary is performed $2^{n-i-1}$ times and the bits from all the previous circuits are received through the \texttt{remote\_c\_if} method in lines 15-17. Then, a Hadamard gate is performed and the bit is measured and transmitted to all subsequent circuits through the \texttt{measure\_and\_send} method in lines 22-23. The flow of information between the circuits in different QPUs is depicted in Figure \ref{subfig:IPEA_qpus}. Finally, note that the application of several classical controlled gates on each subcircuit is not optimal. The optimal way would be to receive all classical bits of previous circuits, then computing the rotation angle from them and finally apply a unique rotation gate. As this flow is not yet supported on CUNQA, the first but suboptimal case was implemented.

\subsection{Quantum communications: distributed QPE}
\label{subsec:quantum-comm}

The QPE has two quantum registers: the ancilla qubits and the qubits where the controlled $U^{2^t}$ are applied. To  execute each part in a different QPU, the controlled $U$ gates will be transformed into distributed gates. The telegate and teledata procotols allow the execution of a controlled gate with the control and the target qubits being from different QPUs. In the use case here presented, the telegate protocol is the one employed but the algorithm could be also distributed through the teledata one. 

The implementation of this case can be seen in the code of Figure~\ref{code:dist_QPE}. In lines 1 to 18, the construction of the circuits corresponding to each quantum register is performed. Specifically in line 11, the expose method \cite{NetQIR} can be seen in use. This is the one implemented the telegate protocol, as explained in Section~\ref{subsec:cunqa_api}, and the $CR_z$---which is the $U^{2^t}$ gate in this particular case---is the one gate added to the protocol. The rest of the code, from line 21-25, sends the circuits to execute in the same manner than the classical-communication model with the employment of the \texttt{run\_distributed} method.

\begin{figure}[!t]
    \centering
\begin{lstlisting}[style=mypython]
    # Defining the distributed QPE circuits
    ancilla_circuit = CunqaCircuit(num_ancilla_qubits, id = "ancilla_circuit")
    register_circuit = CunqaCircuit(1, id = "register_circuit")

    # Initialization of the eigenstate |1>
    register_circuit.x(0) 
    
    for i in range(num_ancilla_qubits):
        ancilla_circuit.h(i)
        
    # Performing the controlled unitary through telegate
    for i in range(n_ancilla_qubits): 
        with ancilla_circuit.expose(num_ancilla_qubits - 1 - i, register_circuit) as rcontrol:
            theta =  2 * pi * 2**(i + 1) * phase # pi = 3,141592...
            register_circuit.crz(theta, rcontrol, 0)
            
    # Inverse QFT (not available in the API, abbreviated for clarity)
    ancilla_circuit.IQFT(0, num_ancilla_qubits) 
    ancilla_circuit.measure_all()
    
    circuits = [ancilla_circuit, register_circuit]

    # Getting available vPQUs
    qpus_QPE  = get_QPUs()

    # Submitting the jobs and gathering results
    distributed_jobs = run_distributed(circuits, qpus_QPE)
    results = gather(distributed_jobs)
\end{lstlisting}
    \caption{Distributed QPE with quantum communications code.}
    \label{code:dist_QPE}
    \Description[Distributed QPE code.]{Python file designing the distributed QPE circuit with quantum communication and executing it using CUNQA}
\end{figure}

\subsection{Obtained results}
\label{subsec:results}

The purpose of this section is to discuss the results obtained for the three DQC models above in order to gain a better understanding of the platform. In all cases the phase to be estimated was $\phi = 1/\pi$, AerSimulator was used as the simulator, and each vQPU was deployed on a CESGA HPC node belonging to the QMIO classical cluster~\cite{qmio_user_guide}. Each vQPU was equipped with the same classical resources: 4 cores and 15GB of memory per core. As a base reference the case of no communications with one vQPU and $10^{8}$ shots will be used. This number of shots far exceeds what is necessary for estimating the phase---for instance the quantum-communication case uses $10^{4}$ shots with the same result---yet it was employed to better showcase the time speedup.

\begin{figure}[!t]
    \centering
    \includegraphics[width=0.5\linewidth]{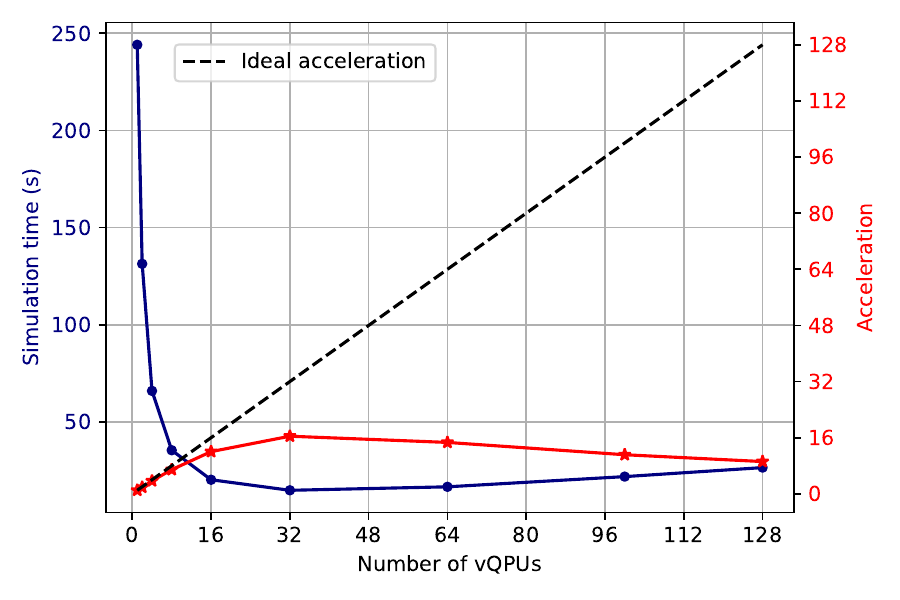}
    \caption{Simulation time (navy blue) and acceleration (red) for estimating $\phi = 1/\pi$ with the QPE algorithm with $n=16$ ancilla qubits.}
    \label{fig:no-comm_results}
    \Description[Graph of the no-communication case]{Graph showing the simulation time and efficiency of the no-communication case}
\end{figure}

In the no-communication case, a range of vQPUs up to 100 was employed, all of them with $n= 16$ ancilla qubits. Figure \ref{fig:no-comm_results} illustrates how simulation time decreases with the increment of the number of vQPUs used in the parallelization (navy blue dots, left axis). From 32 vQPUs the time starts growing due to the overhead introduced by the distribution and gathering of the information, making making the increase in the number of vQPUs not profitable. This is also reflected on the acceleration curve (red stars, right axis), that represents the quotient $\textnormal{time for 1 vQPU}/\textnormal{time for n vQPUs}$. At the beginning, that curve aligns with the ideal case, but, once overhead takes on, acceleration decreases. The 100 vQPU case is emphasized in Table~\ref{tab:exec_comparison} because the number of shots per vQPU is reduced by two orders of magnitude to $10^{6}$. This is the maximum number of shots that has been tested for classical communications due to limitations in execution time.

In the classical-communication case, IPEA employs as many subcircuits---depicted in Figure~\ref{subfig:IPEA_block}---as ancilla qubits in the base QPE, which in this context is 16. Consequently, the natural number of vQPUs required matches the number of subcircuits. Table~\ref{tab:exec_comparison} shows that the simulation time in this case improves upon the base scenario, though it does not reach the performance achieved by the no-communication case with 100 vQPUs, even though the latter is not the optimal one. This difference arises from two factors: first, as discussed in Section~\ref{subsec:virtual_qpus}, in the no-communication model the circuit is handed over to the simulator as a black box, allowing the simulator to apply its own internal optimizations. In contrast, in the classical-communication model the extraction and modification of gate execution from the simulator removes potential optimizations. Second, the inclusion of classical communication directives introduces inherent synchronization delays, which are absent in standard quantum gate execution and thus contribute to longer simulation times.

In the quantum-communication case, only two vQPUs were employed, as discussed in Section~\ref{subsec:quantum-comm}. As shown in Table~\ref{tab:exec_comparison}, the quantum-communication model exhibits execution times two orders of magnitude greater than those of the base case. Increasing the number of vQPUs would further degrade performance, as it would also increase the number of circuit gates. This occurs because distributing additional controlled gates requires the application of a distribution protocol, which introduces a fixed number of extra gates per operation. This, combined with the single-process simulation, implies the simulation time scales proportionally with the number of vQPUs involved.

Finally, the estimated phase was the same for the three DQC schemes and equal to
\begin{center}
    $\phi \simeq 0.\underline{318313}5986328125\pm 0.000015$,
\end{center}
which is consistent with the theoretical $1/\pi = 0.\underline{318309}886183790...$ .
This is the main result of this work: CUNQA effectively emulates DQC algorithms under the three communication schemes.

\begin{table}
    \begin{tabular}{cccccr}
    \toprule
    & nº vQPUs & Shots/vQPU & Phase ($\phi$) & Estimated phase & Execution time (s)\\
    \midrule
    & $1$ & $10^{8}$ & $1/\pi$ & $0.3183135986328125$ & 291.366 \\
    No comm. & & & &  \\
    & $100$ & $10^{6}$ & $1/\pi$ & $0.3183135986328125$ & $16.641$ \\
    \midrule
    Classical comm. & $16$ & $10^{6}$ & $1/\pi$ & $0.3183135986328125$ & $77.760$ \\
    \midrule
    Quantum comm. & $2$ & $10^{4}$ & $1/\pi$ & $0.3183135986328125$ & $13974.584$ \\
    \bottomrule
    \end{tabular}
\caption{QPE results using the three DQC schemes.}
\label{tab:exec_comparison}
\end{table}

%% file: capitulos/4_related_work.tex
\section{Related work}
\label{sec:related_work}
Several tools have been developed aiming for integrating quantum computing in HPC. It is important to, now that CUNQA has been thoroughly explained and contextualized, explain the differences with these tools in order to understand the novelty of CUNQA.

\begin{itemize}
    \item Quantum Framework (QFw)~\cite{chundury2024qfw, beck2024, shehata2024frameworkintegratingquantumsimulation, shehata2026}. The QFw tries to mimic a GPU accelerator stack, as explained in~\cite{shehata2026}. In this software stack they define two critical components: the Quantum Programming Interface (QPI) and the Quantum Platform Manager (QPM). The QPI mediates the application layer access to the middleware, where it can retrieve information about the resources available, insert compilation passes for the toolchain and more. More generally, it simplifies the customization and the interaction with the lower parts of the abstraction model, however, the QPI does not define a programming model---as Qiskit, for instance, does. On the other hand, the QPM, serves as a sort of driver, emerging as the responsible for communicating with the quantum device. It defines an API such that the QPI can interact with every quantum device indistinctively, although this is a current area of research for them.    
    \item Munich Quantum Software Stack (MQSS)~\cite{kaya2024, burgholzer2025mqss}. The MQSS software stack is very similar to QFw, to the point that they even have a so called QPI~\cite{kaya2024qpi}, even though this QPI does define a programming model in C, removing the need of using an external programming language. Moreover, they also define an API that abstracts away the specifics of the quantum devices, called quantum Quantum Device Management Interface (QDMI), which defines an interface that---as they claim in their documentation---is \textit{heavily inspired by the design of the OpenCL API for parallel programming of heterogeneous systems}. 
    \item A Co-Execution Environment for Quantum and Classical Resources (CONQURE)~\cite{mahesh2025conqure}. This work developes two tools: CONQURE cloud, which they claim is an \textit{interoperable alternative for AWS’ cloud queue for quantum devices}; and OpenMP-Q, which extends QpenMP-Q by leveraging its current support for accelerators with the aim of modifying quantum task execution in HPC. These two ways of operating with CONQURE make it a very versatile option, although it has some disadvantages, such as OpenMP-Q model's reliance on a Python script as an intermediate step for execution.
    \item QCCP~\cite{Du2025}. This tool defines a programming model that aims to \textit{shield the super-heterogeneous} nature of HPC environments nowadays, representing a development in the application layer of Figure~\ref{fig:software_stack}. This work is completely complementary to the QFw, in which they did not care about the programming model employed. 
    \item Even though it is not a tool, it is also worth mentioning the work executed by Claudino et al.~\cite{claudino2024}, in which they employed the eXtreme-scale Accelerator programming framework (XACC)\cite{mccaskey2020xacc} in combination to HPC resources to simulate, in the case of their article, the multi-contracted variant of the VQE algorithm (MC-VQE). This work showed that even though XACC is not thought to be an HPC tool explicitly, it can be used in such ways.
\end{itemize}

CUNQA distinguishes itself from all other tools in one essential aspect: it treats DQC as a central element of its design, assigning it the same level of importance as integration within HPC environments. It is, to the best of our knowledge, the only tool that explicitly addresses the emulation of DQC architectures in HPC contexts, and specifically, the first to confront the challenge of simulating quantum communication between emulated QPUs.

Structurally and conceptually, CUNQA shares similarities with QFw and MQSS. However, these works place strong emphasis on the middleware layer within their infrastructures, whereas, as previously discussed, middleware is not a primary concern in CUNQA. This distinction stems from their respective goals: those tools aim to consolidate existing developments in quantum computing and adapt them for HPC, while CUNQA adopts an exploratory approach focused on the future role of quantum computing within HPC, placing DQC at the core of its design. CONQURE, by contrast, provides a cloud-based platform and a modified version of OpenMP capable of supporting quantum accelerators. Both directions diverge from that of CUNQA: the former because a cloud-based system is incompatible with the notion of an accelerator, and the latter because it seeks to adapt a classical framework to the quantum workflow, rather than designing a quantum-oriented approach from the outset. Lastly, QCCP differs fundamentally from CUNQA, as it proposes an entirely new programming model, whereas CUNQA seeks to leverage the most widely adopted model possible to facilitate user adaptation and integration with existing workflows (that is why \texttt{CunqaCircuit} is so similar in use to Qiskit's \texttt{QuantumCircuit}).

%% file: capitulos/5_future_work.tex
\section{Future work}
\label{sec:future_work}

The continuous improvement of CUNQA and the implementation of new features requested by users is a sure bet for the future of the tool. Steps in this direction could range from adding new simulators, to improving the efficiency, quality, and design of the application code, to implementing high-level external libraries that exploit CUNQA as a simulation infrastructure.

One of the ways to this improvement would be the implementation of the dynamic execution of gates without the concept of circuits: following the path set by IBM~\cite{IBM_roadmap}, in a few years there will be quantum computers that support circuits with such a large number of gates that the encoding of the circuits would be on the order of hundreds of gigabytes, an obvious problem when having to move such a large amount of data. Anticipating this problem, CUNQA instructions---gates and measurements---would be sent to the vQPU from the client in small blocks of instructions so that the vQPU will update its status as instructions arrive. A final instruction would inform the vQPU that there are no more instructions to execute, opening up two possibilities depending on the characteristics of the vQPU and the instructions sent: the final status before the measurements could be used to sample the necessary shots, or the measurements could be applied like any other instruction and the sending of instructions would start again until all the shots were executed. Furthermore, to optimize the transmission of these instructions, they would be sent in binary format (using a binary converter developed at CUNQA) and would be applied directly to the vQPU via a hash of binary instructions to instructions supported by the vQPU simulator. This conversion to binary not only improves the efficiency of sending instructions through a traditional communications channel, but also completely eliminates the need for intermediate representations of circuits and instructions. 

This does not imply ending support for the classical circuit execution style. As hinted in Section~\ref{subsec:middleware}, adopting a more efficient IR and expanding middleware functionality are viable paths for future improvement. Using JSON as the IR constrains optimization and poses a memory-efficiency problem compared with representations that avoid string literals.

Regarding the currently supported simulators, they will be updated to their last version. Moreover, as stated in Section~\ref{subsec:virtual_qpus}, in the version of CUNQA presented in this work only the no-communication model supports noise simulation. A future improvement is to add noise support to the classical- and quantum-communication models. Beyond simulating the noise of the standard gates supported by the simulator, this will require simulating the noise of the classical and quantum communication directives, which is a challenge in its own.

Finally, CUNQA has so far been installed and tested at a single HPC center: Galicia Supercomputing Center (CESGA). A natural next step is deployment at other HPC centers to test robustness and expand the user base. This will, in turn, likely lead to additional improvement requests from its users.

%% file: capitulos/6_conclusions.tex
\section{Conclusions}
\label{sec:conclusions}

CUNQA was presented as, to the best of our knowledge, the first tool capable of emulating the three DQC schemes in HPC environments. To enable this emulation, the concept of a virtual QPU (vQPU) was introduced: a classical process running in an HPC environment that simulates quantum tasks as if it were a real QPU. These vQPUs constitute the foundational blocks of the DQC models.

All three DQC models were defined and implemented in this work. The no-communication scheme is a classical distribution of quantum tasks across vQPUs; in this scheme---also called embarrassingly parallel---the vQPUs do not share information at runtime. The classical-communication scheme likewise distributes quantum tasks classically, but includes MPI-like instructions that synchronize vQPUs. Specifically, a measurement result in one vQPU is sent to another vQPU that awaits it to classically control a quantum operation. Finally, the quantum-communication scheme is a purely quantum distribution case, incorporating quantum-communication protocols into the tasks. Its implementation requires a single process to simulate quantum tasks that share quantum information, in contrast to the other two models where each vQPU is responsible of simulating its own task. This architectural change negatively impacts simulation time, making the emulation of quantum-communication models not achieving improvement in times.

From a software perspective, CUNQA is organized into three layers. The first is a low-level layer that implements DQC emulation in C++ and comprises the vQPUs and their associated components. The second layer is a Python interface that enables the definition and execution of distributed circuits under the three DQC schemes. It is intended to remain applicable when vQPUs are replaced by real QPUs with any form of inter-QPU communication. The third layer represents the middleware, which has the goal of connecting the other two. Each layer is designed to minimize dependencies and decouple CUNQA from third-party libraries.

In conclusion, CUNQA is a step forward in the growing field of DQC. It is designed for testing the novel architectures this field presents in order to better understand how to work with them and, specially, how to program its problems. Ongoing development and adoption of emerging techniques are core goals, with the aim of remaining relevant amid the rapid evolution of quantum computing.

%% file: capitulos/7_data_availability.tex
\section{Data availability}
\label{sec:data_availability}

CUNQA is an open-source project and its code can be found in the GitHub repository \url{https://github.com/CESGA-Quantum-Spain/cunqa}. The code version presented in this work is 1.0.0.